\documentclass[9pt]{article}

\usepackage[british]{babel}

\usepackage{amsmath}
\usepackage{amsfonts}
\usepackage{tikz}
\usepackage{xfrac}
\usepackage{bbm}
\usepackage[most]{tcolorbox}
\usepackage{amssymb}
\usepackage[utf8]{inputenc}
\usepackage{url}
\usepackage{hyperref}

\usepackage{authblk}
\usepackage{mathptmx}

\definecolor{white}{RGB}{255,255,255}
\definecolor{greyYed}{RGB}{204,204,204}
\definecolor{yellowYed}{RGB}{255,204,0}
\definecolor{greenYed}{RGB}{153,204,0}
\definecolor{orangeM}{RGB}{217, 83, 25}
\definecolor{blueM}{RGB}{0, 114, 189}
\definecolor{yellowM}{RGB}{237, 177, 32}
\definecolor{purpleM}{RGB}{126, 47, 142}
\definecolor{greenM}{RGB}{119, 172, 48}
\definecolor{cyanM}{RGB}{77, 190, 238}
\definecolor{redM}{RGB}{162, 20, 47}

\hypersetup{
    colorlinks=true,
    filecolor=purpleM,
    linkcolor=blueM,
    citecolor=redM,
    urlcolor=greenM
}

\newcommand{\tcbM}[2]{
	\tcbhighmath[
		on line,
		boxsep=2pt,
		left=2pt,right=2pt,top=2pt,bottom=2pt,
		colback=#1!40!,colframe=white,
		highlight math style={enhanced}
	]{#2}
}

\newcommand{\sref}[1]{Section~\ref{#1}}
\newcommand{\tref}[1]{Table~\ref{#1}}
\newcommand{\fref}[1]{Figure~\ref{#1}}
\newcommand{\aref}[1]{Appendix~\ref{#1}}

\newcommand{\toprule}{\hline}
\newcommand{\midrule}{\hline}
\newcommand{\bottomrule}{\hline}

\begin{document}

\title{RSSi-based visitor tracking in museums\\via cascaded AI classifiers and\\coloured graph representations}

\author[a]{\underline{Elia Onofri}}
\author[b]{Alessandro Corbetta}

\affil[a]{\small Department of Mathematics, Roma Tre Univ., Rome, Italy \textsc{eonofri@uniroma3.it}}
\affil[b]{\small Department of Applied Physics, Eindhoven Univ.\ of Tech., Eindhoven, The Netherlands \textsc{{a.corbetta@tue.nl}}}

\date{}

\maketitle

\begin{abstract}
	Individual tracking of museum visitors based on portable radio beacons, an asset for behavioural analyses and comfort/performance improvements, is seeing increasing diffusion.
	Conceptually, this approach enables room-level localisation based on a network of small antennas (thus, without invasive modification of the existent structures).
	The antennas measure the intensity (RSSi) of self-advertising signals broadcasted by beacons individually assigned to the visitors.
	The signal intensity provides a proxy for the distance to the antennas and thus indicative positioning.
	However, RSSi signals are well-known to be noisy, even in ideal conditions (high antenna density, absence of obstacles, absence of crowd, \dots).
	
	In this contribution, we present a method to perform accurate RSSi-based visitor tracking when the density of antennas is relatively low, e.g.\ due to technical constraints imposed by historic buildings.
	We combine an ensemble of ``simple'' localisers, trained based on ground-truth, with an encoding of the museum topology in terms of a total-coloured graph.
	This turns the localisation problem into a cascade process, from large to small scales, in space and in time.
	Our use case is visitors tracking in Galleria Borghese, Rome (Italy), for which our method manages $>96\%$ localisation accuracy, significantly improving on our previous work (J.\ Comput.\ Sci.\ 101357, 2021).
	
\end{abstract}


\section{Introduction}\label{sec:intro}

The behavioural analysis of museums' visitors has a long-standing multidisciplinary tradition~\cite{robinson1928book, FENG2021107329}, and underlies the capacity of profiling exhibitions, increase visitors' comfort and safety, improve public reception, increase the number of sold tickets, and enhance artworks preservation~\cite{yoshimura2014}.
For instance, understanding visitors' dynamics unlocks focused tuning of visiting paths, collocation of pieces, and access/ticketing strategies.
The recording and analysis of individual visitors' trajectories -- possibly across the entire museum venue -- is a great asset towards such behavioural analyses~\cite{JCompSci}, allowing even to re-generate plausible visiting patterns~\cite{balzotti2018, pluchino2014}.
Their feasibility has significantly grown during the last decades thanks, particularly, to the diffusion of Internet-of-Things (IoT) technologies~\cite{falk2016book, yalowitz2009, piccialliMuseums}, which enabled individual tracking needlessly of invasive structural modifications (e.g.\ as happens with overhead optical tracking sensors \cite{submittedCorbettaPed}).
Nowadays, countless portable IoT devices broadcast periodically their identities on the Bluetooth and/or Wi-fi networks.
Measuring at different locations the strength of these signals, the so-called RSSi (i.e.\ the Received Signal Strength indication), yields a mechanism to perform localisation and tracking (cf.\ also reviews \cite{review1Localisation, review2Localisation}).
Such an approach has been employed in different fields, allowing tracking in healthcare facilities~\cite{BLEhospitals} and smart buildings~\cite{IoTSmartBuildings}.

From an operational point of view, one provides (consenting) individuals (and/or groups) with a portable IoT \textit{beacon} (note that personal mobile phones, modulo privacy and randomisation issues, could be used likewise~\cite{yoshimura2014}).
Antennas preemptively deployed measure the RSSi of the periodic advertisements of the beacon.
Thus, the visitor position can be roughly approximated with that of the antenna measuring the highest RSSi value~\cite{piccialli2019a}.
In principle, high antennas densities could also allow precise localisation through signal tri/multi-lateration~\cite{wang2009multilateration}.
On the other hand, even in optimal conditions (e.g.\ line of sight to the beacon, absence of radio interference) RSSi values typically suffer from high fluctuations~\cite{beder2012}.
Additionally, in historic buildings, a frequent location for museums, massive antennas deployments are impossible due to architectural constraints, while room-level tracking allows sufficient insights.
Museums often feature complex geometries rich of obstacles and -- especially in old constructions -- a wide mixture of thick and thin walls with narrow and wide doors.
In combination with crowding, these yield even noisier RSSi signals with a quick decay, causing ambiguous or even void positioning readings.
These constraints jeopardise the success of any approach based on instantaneously ``maximum RSSi'' readings (argMax), even when the ambition is the sole room-level localisation.
To deal with these limitations, we proposed in~\cite{JCompSci} a method based on an end-to-end Multi-Layer Perception (MLP) neural network.
This MLP significantly improved room-level localisation performances over na\"ive argMax policies by considering short RSSi time sequences.
In particular, the localisation (binary) accuracy `acc' (percentage of correct room-level localisation outputs) was increased from 54.7\% (argMax policy) to 85.8\% (MLP), considering an experimental dataset with 900 real-life trajectories collected by the same authors in the Galleria Borghese museum in Rome, Italy.
Analysing signals in short time windows enabled higher robustness against the fluctuations of RSSi.
However, extending the time window still proved insufficient to tackle environments with a coarse/irregular antennas distribution, for which the MLP localisation is close to random.
In general, typical museum geometries are an obstacle to end-to-end learning, including physically disconnected areas, although geometrically close, that can be ambiguously represented by RSSi readings.
It is reasonable thus, that a methodology as an MLP trained on short RSSi windows is incapable of resolving these aspects: no ``expert'' knowledge, such as constraints dictated by museum topology (preventing, e.g., ``nonphysical jumps'' amongst rooms) is injected.

In this work, we propose a methodology to perform RSSi-based visitor tracking within a sparse antennas grid that combines an ensemble of trained classifiers, henceforth referred to as \textit{localisers}, with an expert encoding of the museum topology in terms of a room-scale total-coloured graph.
Since we replace a ``complex'' predictor with multiple simpler ones, which operate at different space- and time- scales, and thanks to the injection of expert knowledge, we manage to significantly increase the localisation accuracy up to $\text{acc}\approx 96.8\%$ for our use case at Galleria Borghese.
Additionally, the total-coloured graph representation of a museum enforces a room-scale metric $\mathcal D$ and, consequently, a (room-level) trajectory-scale metric $\mathcal W$.
We employ such metrics as indicators to compare reconstructed trajectories with the corresponding ground-truth, providing a more insightful performance analysis than the simple binary accuracy.

The rest of the paper is organised as follows: in \sref{sec:IoT}, we review our experimental campaign at Galleria Borghese, the use-case for our methodology, and review the basics of RSSi-based tracking; in \sref{sec:totalColouredMuseum} we introduce a topological representation for museum-like environments, based on total-coloured graphs; these are at the basis of the cascaded localiser method that we introduce in \sref{sec:multiLSTM}.
The method performances are addressed in \sref{sec:results}.
A final discussion in \sref{sec:conclusions} closes the work.

\section{RSSi-based visitor tracking at Galleria Borghese}
\label{sec:IoT}
In this section, we review the basics of RSSi-based tracking considering our measurement installation at the Galleria Borghese (GB) museum, Rome, Italy, which we thoroughly described in our previous work~\cite{JCompSci}.
GB hosts hundreds of art pieces in a relatively small area, and consists of $21$ exhibition volumes divided into two floors, which are accessible via $3$ entrances.
Due to logistic constraints, visits are partitioned in $2$ hour-long slots, non-overlapping.
Like many other cultural heritage sites, GB has heavy limitations when it comes to set up a data-gathering infrastructure, e.g.\ no internet connection and, even worst, a very limited number of electrical outlets to power any sensing infrastructure.
Room-level tracking based RSSi-measurements, demanding at least one available electrical plug per room, is thus a viable way to bypass these constraints.
Tracking at room-level entails (anonymously) associating to each visitor their room-level trajectory $\mathfrak{t}$ satisfying
\begin{equation}\label{eq:trajectory}
	\mathfrak{t} = (\mathfrak{t}_0,\mathfrak{t}_1,\ldots,\mathfrak{t}_{T-1}),
\end{equation}
where $\mathfrak{t}_t$ states the room (within a finite virtual room set $\{R_1, R_2,\ldots, R_n\}$) in which the visitor is located at discrete time $t$, considering a regular time sampling $t\in{0, \ldots, T-1}$.

We employ small commercial Bluetooth-Low-Energy (BLE) beacons that are assigned to each guest at the beginning of the visit and given back at the end.
Each beacon broadcasts a signal carrying its unique identifier which we captured by a fleet of $A = 14$ receiving antennas deployed as good as allowed by the available electrical outlets (ten antennas on the first floor, roughly one per room, and four on the second floor which we consider as a single virtual room).
We employ single-board Raspberry Pi computers as antennas, which are programmed to forward every 5 seconds the measured RSSi information (along with its timestamp) to a remote database.
Such a system proved to be reliable with dozens of beacons broadcasting at the same time within a small area.

\begin{figure}[t]
	\begin{center}
	\includegraphics[width=0.39\linewidth]{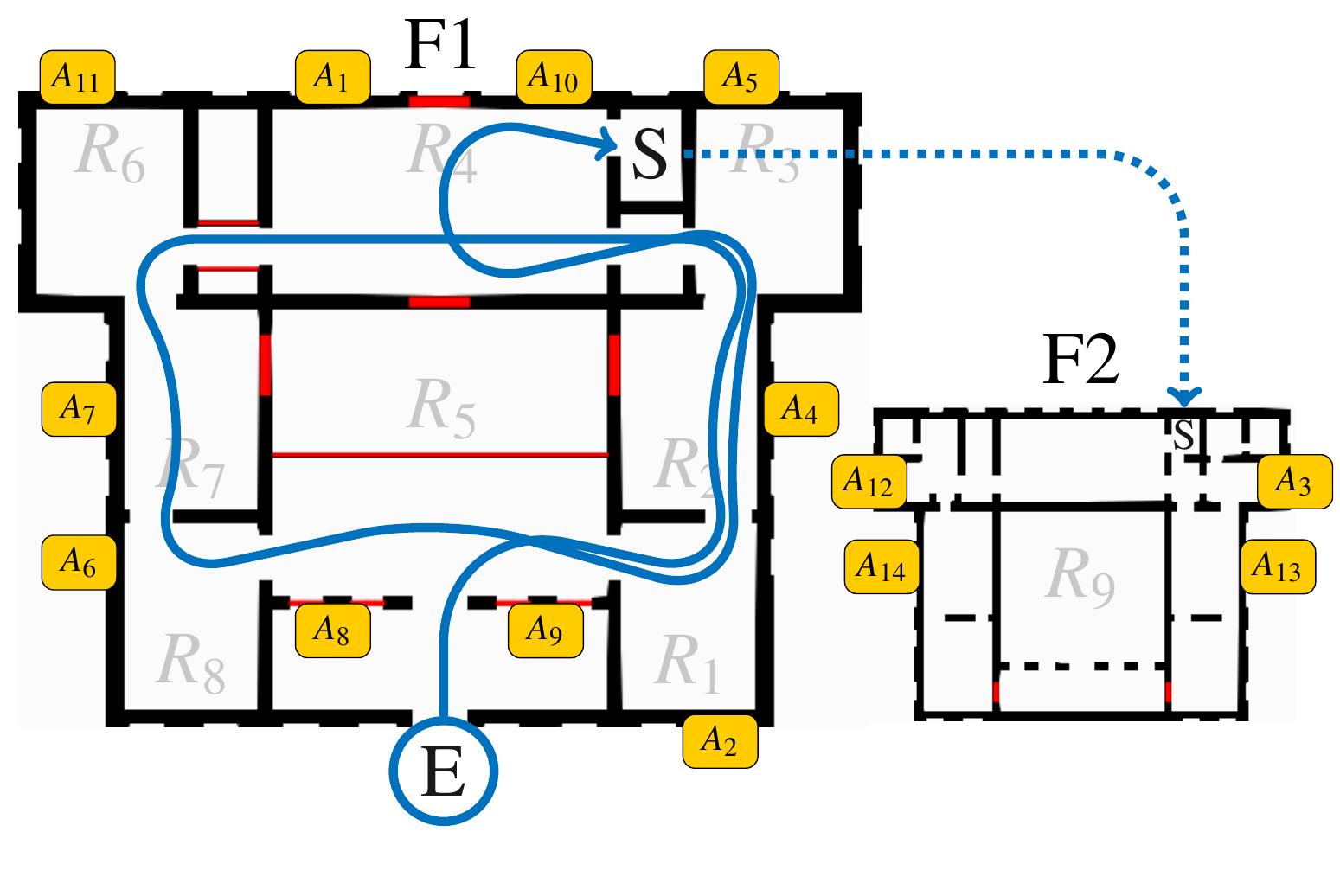}
	\includegraphics[width=0.6\linewidth]{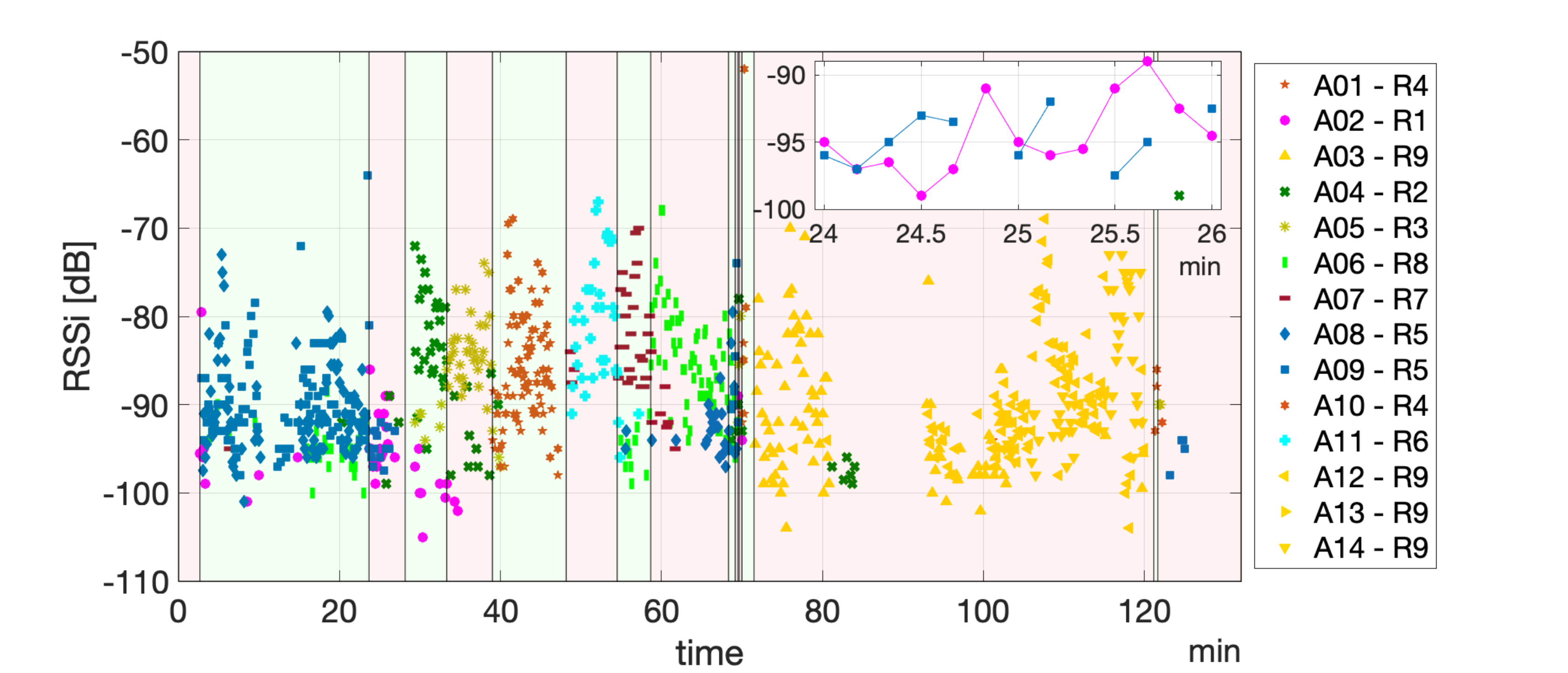}
		\caption{
			\textbf{(a)} Galleria Borghese floor plan.
			We report the position of antennas ($A_{xy}$) and rooms ($R_x$).
			Red lines represent closed doors.
			We include in blue a sketch of the trajectory whose RSSi measurements are in panel \textbf{(b)}.
			The trajectory begins on the first floor (F1) and follows a counter-clockwise path before reaching the second floor (F2).
			\textbf{(b)} RSSi measurements for a real visitor trajectory (sketched in panel \textbf{(a)}).
			The readings from the $A=14$ antennas are coloured depending on the virtual room (thus, measurements taken by any of the antennas on the second floor come with the same colour, yellow triangles).
			Each room change is represented by a different background colour.
	}
		\label{fig:RawBeacon}
	\end{center}
\end{figure}

In \fref{fig:RawBeacon}\textbf{(b)} we report the RSSi readings by the different antennas for the trajectory sketched \fref{fig:RawBeacon}\textbf{(a)}.
With respect to \fref{fig:RawBeacon} we report different critical issues: (i) the signal recovered by adjacent antennas may overlap continuously as highlighted by the inset, (ii) no signal is detected for a long amount of time, especially on the second floor (minutes $85\sim95$), (iii) antennas from the first floor may capture the beacon signals, even when it is on the second floor, (iv) some random antennas may detect beacons even after the visitor left the exhibition area (e.g.\ at the cafeteria or the wardrobe).

The gathered RSSi data does not have uniform temporal sampling so, we resample the signal with a fixed time-stepping $\Delta t=10\,$s.
The choice of $\Delta t$ is a trade-off between signal granularity and the desired resolution.
Our choice yields $6$ samples per minute, which is small enough with respect to the typical permanence and traversal time of a room.
Given the captured RSSi from a single beacon, the output of this resampling procedure is a $A \times T$ matrix $\mathcal R$ (where $T$ is the duration of the visit divided by $\Delta t$).
In~\cite{JCompSci, centorrino2019} we proposed two methodologies to process such a matrix to estimate individual trajectories, the most effective of which was based on an MLP neural network operating on $\mathcal R$ after a row-normalisation (average over time $=0$, standard deviation over time $=1$), resulting in $\text{acc }=85\%$ localisation accuracy.
Specifically, our MLP was 3-layers deep, its input was a two-minutes long symmetric time window (i.e.\ $A\times (6+1+6)$ dimensional) and the localisation output was delivered via a soft-max activation defined over the room set.

In \sref{sec:multiLSTM} we will introduce a new methodology that improves these results by achieving a global mean accuracy over $96\%$ as we include assumptions on the topological structure of the museum.
In the next Section, we present the graph interpretation of the museum at the basis of the method.

\medskip

\section{Room-scale representation of Museums as total-\-co\-lou\-red graphs}
\label{sec:totalColouredMuseum}

We consider here a generic museum divided into multiple floors; usually, a visitor explores one floor at time instead of changing it continuously.
Two kinds of connections arise from this example: same and different floor links.
However, we can think of other different architectural constraints that may influence a visit, e.g., different buildings.
Similarly, we can provide conceptual subdivisions that concern rooms: e.g.\ if the exhibition is organised in different historical time periods, the typical visitor will likely explore a thematic area at time instead of traversing them randomly.

In the following, we formalise this approach by first sketching the museum as a total-coloured graph and then extracting its emerging clustering.
We also provide a metric definition to make useful insights available when it comes to comparing trajectory reconstruction methods.

\subsection*{Extraction of a total-coloured graph}

\begin{figure}\centering
	\includegraphics[width=0.99\linewidth]{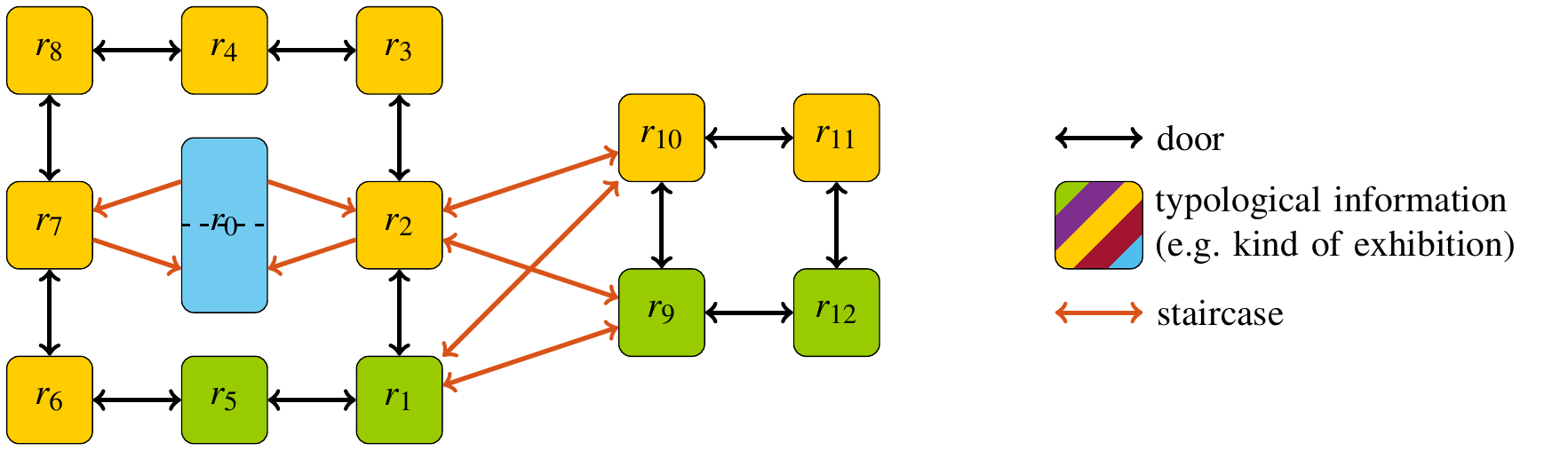}\\
	\textbf{(a)}\\
	\includegraphics[width=0.45\linewidth]{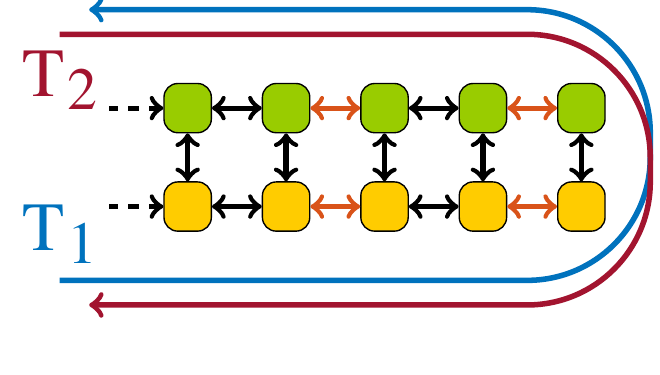}\hfill
	\includegraphics[width=0.45\linewidth]{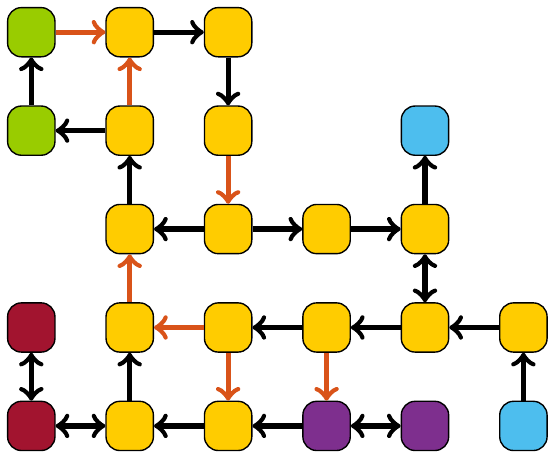}\\
	\textbf{(b)}\hspace{7cm}\textbf{(c)}
	\caption{
    	Use cases for total-coloured graphs in three fictitious museums.
    	\textbf{(a)} A museum inspired by Galleria Borghese (Rome, Italy) with three floors and two kinds of exhibitions.
    	In \sref{sec:totalColouredMuseum} we build our methodology on this example.
    	\textbf{(b)} A segment of a museum where all the three floors host two kinds of exhibitions.
    	Without considering the vertex-colour information, the two trajectories $\textsc t_1$ and $\textsc t_2$ would be very similar, however, they clearly are opposite and therefore different.
    	\textbf{(c)} An example of a directed museum with multiple kinds of exhibitions based on Vatican Museums (Rome, Italy).
    	Entrance and exit artificial rooms are at different locations.
	}
	\label{fig:museumExample}
\end{figure}

In this paragraph we provide, by means of graph theory \cite{DiestelGT}, a natural formalisation of a museum as a graph equipped with different colouring to represent its architectural and conceptual constraints (cf.\ \fref{fig:museumExample}).

A graph is a mathematical structure made of two sets, $V$ and $E$, representing, respectively, entities and connections between them.
Such entities are often called vertices and, in our representation, they symbolise the museum (virtual) rooms $V = \{r_1, r_2, \dots, r_n\}$ (i.e.\ single rooms or aggregates thereof depending on the antenna density).

We formalise rooms connections through ordered couples of rooms $e = (r_i, r_j)$, often called edges.
In regular museums, connections are often bi-directional, meaning both $(r_i, r_j)$ and $(r_j, r_i)$ belongs to the edge set $E = \{e_1, e_2, \dots, e_m\}$; however, also one-way connections may occur (cf.\ \fref{fig:museumExample}\textbf{(c)}).

We enforce a conceptual subdivision of the rooms by means of a set of \emph{typological} information or, in graph theory jargon, colours.
Similarly, we specify architectural constraints, e.g.\ being a door or a staircase, equipping also edges with a colour.
More formally, a vertex-colouring (resp.\ edge-colouring) is a function $\gamma$ that maps $v \in V$ (resp.\ $e \in E$) in a set of natural numbers $C \subset \mathbb N$.
A graph $(V, E, \gamma)$ in which both vertices and edges are supplied with colours is called \emph{total-coloured} graph.

We add additional artificial rooms, marked with a dedicated colour, to represent the entrances/exits of the museum.
These rooms will serve in the trajectory reconstruction process as placeholders for the visitors/beacon before the visit begins, after it ends, and, possibly, whenever no signal is detected.
Conceptually, these rooms shall be sources or sinks of the graph, namely nodes with no incoming or outgoing edges, respectively.

\begin{figure}\centering
	\includegraphics[width=0.30\linewidth]{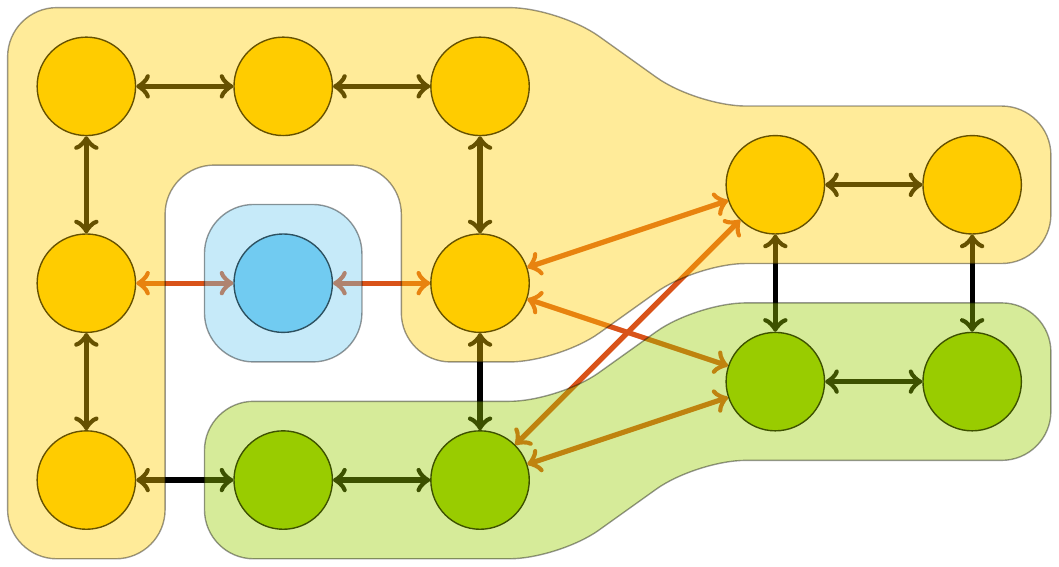}\hfill
	\includegraphics[width=0.30\linewidth]{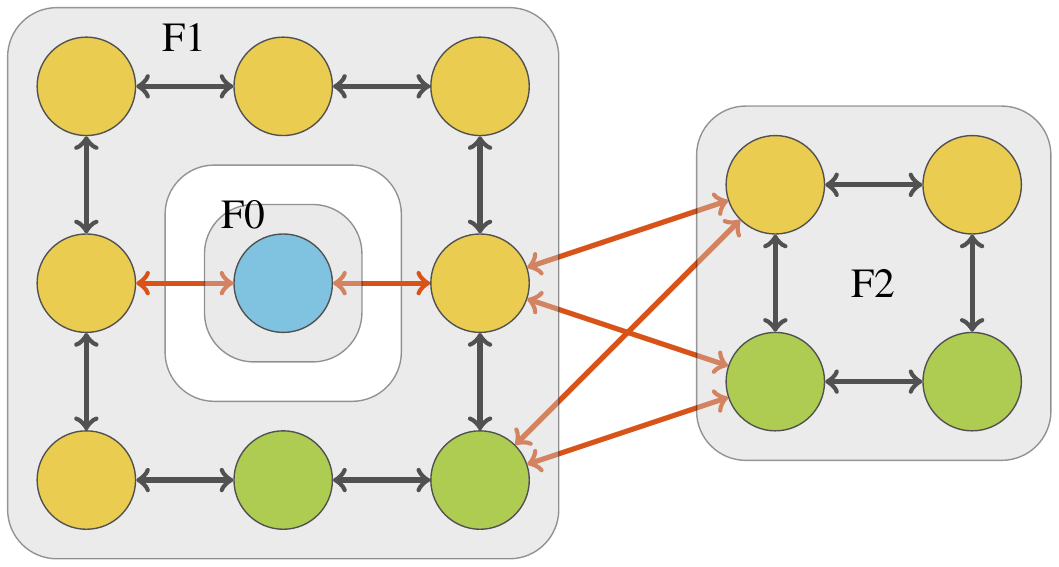}\hfill
	\includegraphics[width=0.30\linewidth]{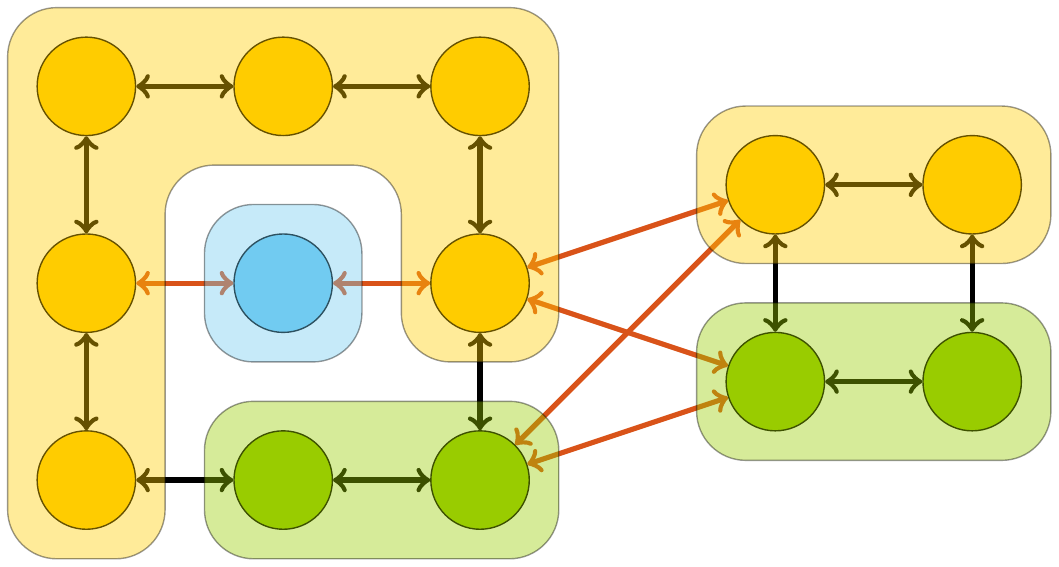}\\
	\textbf{(a)}\hspace{4cm}\textbf{(b)}\hspace{4cm}\textbf{(c)}
	\caption{
		Colour-contractions applied to the graph representation in \fref{fig:museumExample}\textbf{(a)}.
		\textbf{(a)} vertex-colouring: the two kinds of exhibitions join in one single cluster each, suggesting that visitors explore them independently.
		\textbf{(b)} edge-colouring: by contracting over door edges $(\{\leftrightarrow\})$ we obtain a floor-based representation, prompting the idea that visitors would rarely deal with stairs if it is not needed.
		\textbf{(c)} total-colouring: contracting over both vertex-colouring and edge-colouring summaries the hints from the clustering \textbf{(a)} and \textbf{(b)}, detecting four areas, likely visited independently.
	}
	\label{fig:museumContraction}
\end{figure}

\subsection*{Room clustering via colour-contraction}

We now consider \emph{colour-contraction}, a methodology to extract different kind of clustering, i.e.\ partitions\footnote{A partition of a set $V$ is a set of disjoint subsets $K_1, K_2, \dots \subseteq V$ such that their union gives the entire set.} $(K_1, K_2, \dots)$ of the room set $V$, from a coloured graph.
This methodology, originally designed as a fast method to rescale big sets of data~\cite{mansour1993graph}, allows us to infer insights from the typological information stored as colours in vertices and edges.

Depending on which colouring we rely on, we can build different partitions of $V$ by joining two rooms $r_i$ and $r_j$ in the same cluster as follows:
\begin{itemize}
	\item[] \emph{vertex-colouring}:
		if $r_i$ and $r_j$ are connected and share the same colour, then are in the same cluster, say $K_l$, that is:
		\begin{equation}
			\text{if} \quad (r_i, r_j) \in E \quad \text{and} \quad \gamma(r_i) = \gamma(r_j) \quad \text{then} \quad r_i, r_j \in K_l \ .
		\end{equation}
		Note that two rooms do not need to be directly connected to belong to the same cluster.
		This aggregates all the rooms sharing the same conceptual information which are reachable one from another (cf.\ \fref{fig:museumContraction} \textbf{(a)}).
		
	\item[] \emph{edge-colouring}:
		if $r_i$ and $r_j$ are connected by an edge of some specific colour $C' \subsetneq C$, then are in the same cluster, say $K_l$, that is:
			\begin{equation}
			\text{if} \quad e = (r_i, r_j) \in E \quad \text{and} \quad \gamma(e) \in C' \quad \text{then} \quad r_i, r_j \in K_l \ .
		\end{equation}
		As an example, by selecting door connections, this technique contracts all the rooms within the same floor of the museum in single clusters (cf.\ \fref{fig:museumContraction} \textbf{(b)}).
\end{itemize}

\noindent We can also combine these two methodologies, therefore obtaining a clustering of the rooms based on both architectural and conceptual constraints (cf.\ \fref{fig:museumContraction} \textbf{(c)}).

\subsection*{Trajectory and room metric enforced by the total-coloured graph}
To evaluate the quality of our trajectory reconstruction process (to be discussed in \sref{sec:multiLSTM}), we shall rely on a distance between the reconstructed trajectory and the corresponding ground-truth.
A total-coloured graph induces a room-level metric $\mathcal D$ and, on this basis, a trajectory-level metric $\mathcal W$, which we will use for evaluation purposes.

A ``vanilla'' distance between two rooms/vertices for a colourless graph is defined by the least number of edges traversed to reach one room from the other.
Determining such a distance is a well-known graph theory problem, dubbed the \textit{shortest path problem}, which naturally extends to include edges weighing a non-unit distance (i.e.\ weighted edges)~\cite{TarjanShortestPath}.

In the setting of coloured graphs, the colours themselves can be interpreted as weights, e.g.\ the weight of a colour can be proportional to the time employed to traverse that specific kind of edge.
Therefore, the assignment of a weight $w(\gamma(\cdot))$ to each edge-colour enforces also a distance notion between rooms\footnote{\label{fn:discountFactor}As an additional feature, it can be useful to reduce the cost of the first door transition, actually decreasing the weight of short transitions that can occur if visitors stand still by the entrance door.}.

Whenever two rooms come in different colours, since their typological information differs (cf.\ \fref{fig:museumExample}\textbf{(b)}), a penalty $\beta$ is applied to their mutual distance.
Note, whenever $\beta > 0$, the metric $\mathcal D$ does not satisfy the relaxation principle~\cite{Cormen} (which can be still applied by adding $\beta$ \emph{a posteriori}).

The distance $\mathcal D$ is discrete in the vertices of the graph, therefore it can be represented as a $n \times n$ matrix.
The distance is commutative if and only if all the connections in the underlying graph are bi-directional, therefore no assurance can be made \emph{a priori} on the symmetry of $\mathcal D$.
Formally, we have:
\begin{equation}
	\mathcal D_{i,j} \quad =\quad \mathcal D(r_i, r_j) \quad = \quad \beta \cdot \mathbbm{1}_{\{\gamma(r_i) \ne \gamma(r_j)\}} \quad + \quad \sum_{\substack{e \in \Gamma\\\Gamma \text{ shortest path}\\\text{between\ } r_i \text{ and\ } r_j}} w(\gamma(e)) \ ,
\end{equation}
where $\mathbbm{1}_{\{\epsilon\}}$ is the indicator of the event $\epsilon$, i.e.\ $\mathbbm{1}_{\{\gamma(r_i) \ne \gamma(r_j)\}} = 1$ if $\gamma(r_i) \ne \gamma(r_j)$ and $0$ otherwise.

Given two room-level trajectories $\mathfrak{s}$ and $\mathfrak{t}$ spanning $T$ discrete time instants (cf.\ Eq.~\eqref{eq:trajectory}), we can extend the room-level metric $\mathcal D$ to a trajectory-level metric $\mathcal W$ as the component-wise sum of the distances between instantaneous location.
In formulas, it holds
\begin{equation}
	\mathcal W(\mathfrak{s}, \mathfrak{t}) = \sum_{t=0}^{T-1} \mathcal D(\mathfrak{s}_t, \mathfrak{t}_t)\ ,
\end{equation}
where $\mathfrak{s}_t$ and $\mathfrak{t}_t$ are the rooms in which trajectories are at time $t$.
It is assumed that $\mathfrak{s}$ and $\mathfrak{t}$ have the same length $T$.
In an applied perspective, we need to compare also trajectories of different lengths~\cite{JCompSci}.
To this aim, we pad the end of the shorter trajectory with a virtual exit room such as its length matches the longer trajectory.

\section{Cascaded trajectory reconstruction based on colour-clustering}
\label{sec:multiLSTM}

In this section, we show how the multiple clustering of museum rooms we obtained from the total-coloured graph representation lay the foundation for cascaded classifiers that improve room-level localisation accuracy over the methods reviewed in \sref{sec:IoT}.

Our clustering creates a partition of $V$ in aggregates of rooms $K_1, K_2, \ldots$ (clusters) that are,  by construction, explored by visitors one at a time.
Thus, our localisation approach first identifies the visitor in one amongst the different room clusters, say $K_l$, and, only afterwards, in a specific room within $K_l$.
From a practical point of view, this corresponds to building one ``large-scale'' localiser that classifies at cluster-level $K_1, K_2, \ldots$, and multiple ``fine-scale'' localisers, each operating within a single cluster, and returning a specific room.
Each localiser, at either large- or small- scale, operates on all or a subset of $A' \leq A$ RSSi signals, possibly down-sampled in time.
As RSSi are highly fluctuating, analysing data from a time window of, say, $T' \leq T$ samples, instead of the single measurements is mandatory.
We refer to the time window as symmetric if data from both the future and the past is used, and asymmetric otherwise (necessary for online trajectory reconstruction).
In the case of the ``large-scale'' classifier, considering a time window including a down-sampled version of the RSSi signal yielded better performance since it limited the input dimension of the classifier and prevented noisy inputs.

\begin{figure}
	\centering
	\includegraphics[width=0.5\linewidth]{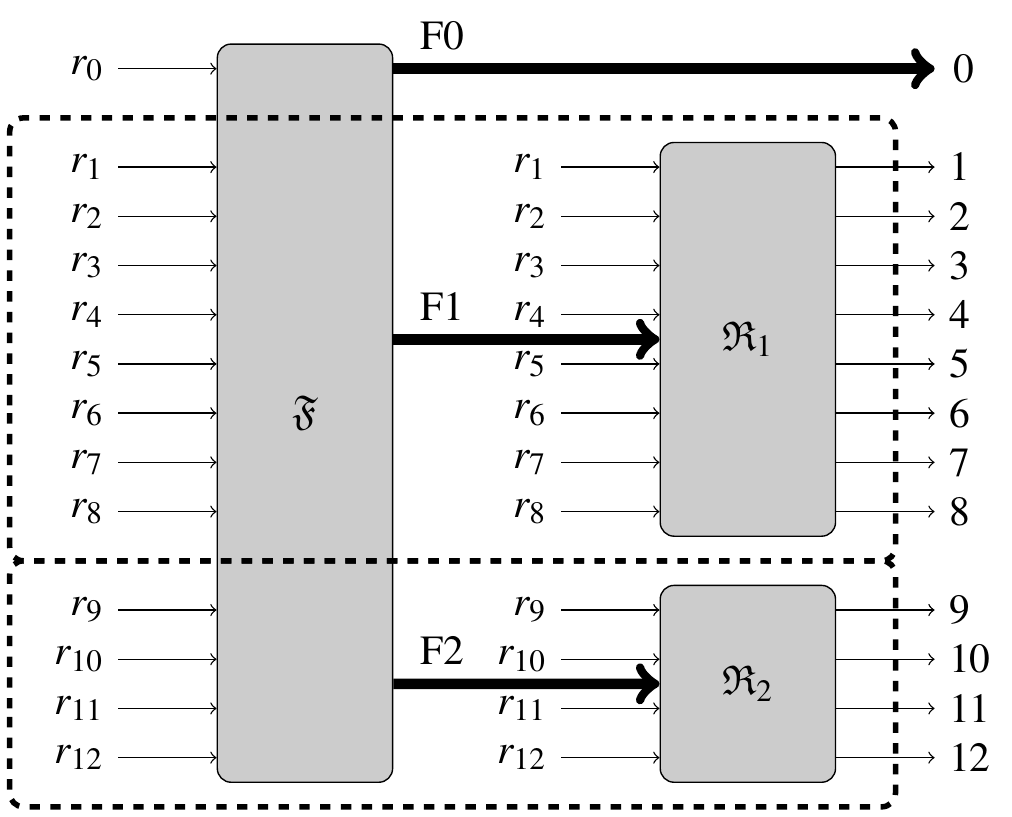}
	\caption{A possible design for a two-layer cascaded selector built upon the museum in \fref{fig:museumExample}\textbf{(a)}, as enforced by the clustering in \fref{fig:museumContraction}\textbf{(b)}.}
	\label{fig:selectors}
\end{figure}

Considering the example in \fref{fig:museumContraction}\textbf{(b)}, we would employ three localisers: a floor selector $\mathfrak F$ fed with the signal from all the antennas that outputs one amongst the three areas (F$0$, F$1$ or F$2$) of the museum; two room selectors $\mathfrak R_1$ and $\mathfrak R_2$ (to be mutually employed depending on the output of $\mathfrak F$) that use the signals from the receivers on their respective floor to output the most probable room within that floor (note that the area F$0$ has a single room associated to, so it does not need a selector $\mathfrak R_0$; cf.\ \fref{fig:selectors}).

\subsection*{Individual localisers}
We now briefly outline different choices for individual localisers.
Depending on the amount of data available one could opt for heuristics or data-driven methods.

Heuristic algorithms heavily depend on the museum topology.
Convolutional kernel approaches (ConvKer, for brevity) yield a baseline.
Specifically, we consider the argMax amongst the antennas, applied to a weighted mean of the RSSi signals within the time window.

Many options are available for classification algorithms trained with ground-truth data~\cite{deeplearningbook}.
We specifically consider the following three, widely used in the literature:
\begin{description}
	\item[] \textit{Multi-Layer Perceptron (MLP).} An MLP~\cite{MacKay} is an artificial neural network made of an ensemble of units (perceptrons) organised in at least three consecutive layers as a directed graph.
	Data is assigned to the first layer units (which are $A' \cdot T'$) and flows into the next layers as a non-linear activation function (sigmoid) applied to a linear combination of the values of the previous layer.
	The last layer provides a classification output as a probability over $A'$ classes via a soft-max function.
	The training procedure iteratively back-propagates through the layers modifying the linear combination weights.
	
	\item[] \textit{Long-Short Term Memory (LSTM).} A LSTM~\cite{reviewLSTM} is a particular kind of recurrent neural network (RNN), i.e.\ artificial neural networks where data can also propagate backwards, designed specifically for time series.
	LSTM are fed with data from the $A'$ sensors time bin by time bin and progressively fill up an internal cell state whose of finite size.
	The data flow from the input and the state to the output, being controlled by multiplicative and non-linear gates.
	The training procedure back-propagates through time in the cell state, modifying the gates weights.
	
	\item[] \textit{Random Forest (RF).} A RF~\cite{reviewRF} is a predictor consisting of an ensemble of randomised decision trees, i.e.\ weaker tree classifiers where each internal node represents a query over the data features and each leaf is a classification label.
	The RF classifies following the majority vote over the tree ensemble, following the concept that weaker classifiers trained on different subsets of data learn various aspects from the features.
	The training procedure uses different dataset bootstrapping in order to train deterministically the various trees individually.
\end{description}

\section{Trajectory reconstruction results in Galleria Borghese}\label{sec:results}

\begin{figure}
	\centering
	\includegraphics[width=0.25\linewidth]{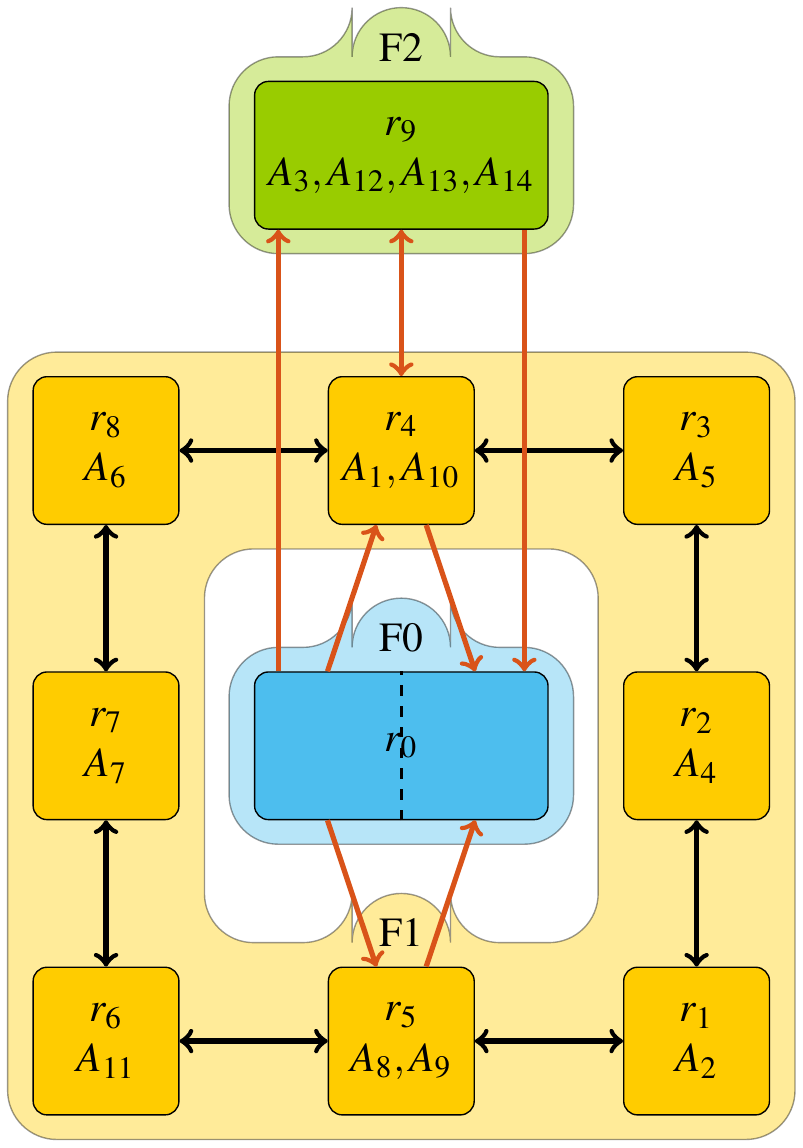}
	\hfill
	\includegraphics[width=0.3\linewidth]{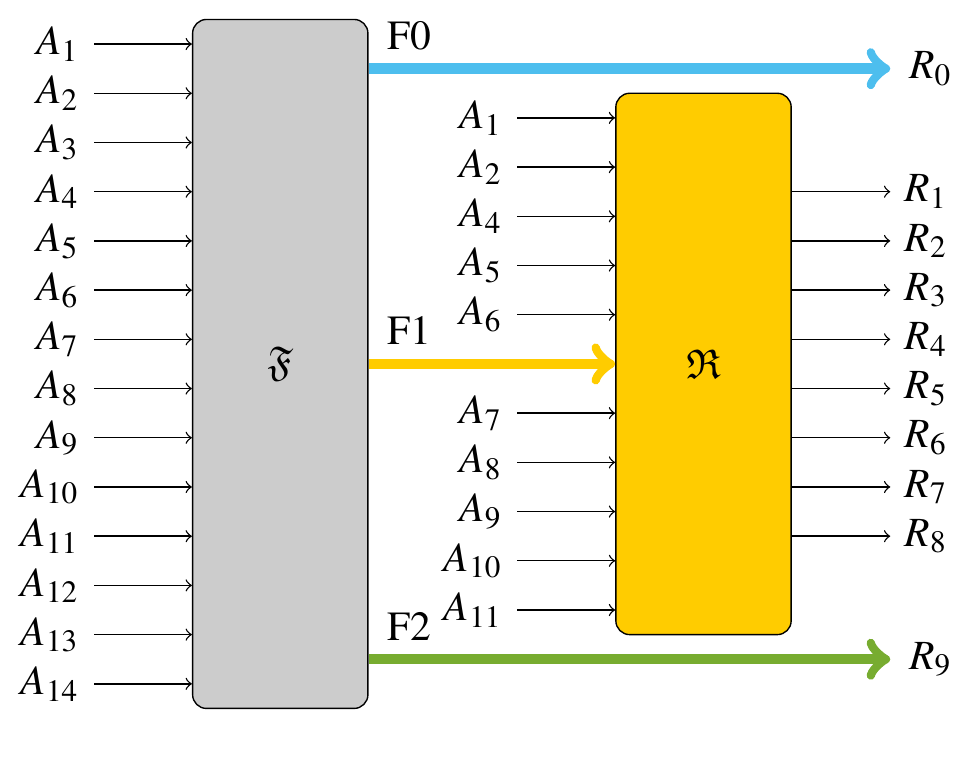}
	\hfill
	\includegraphics[width=0.3\linewidth,trim=2.3cm 7.5cm 3.2cm 8.5cm, clip=true]{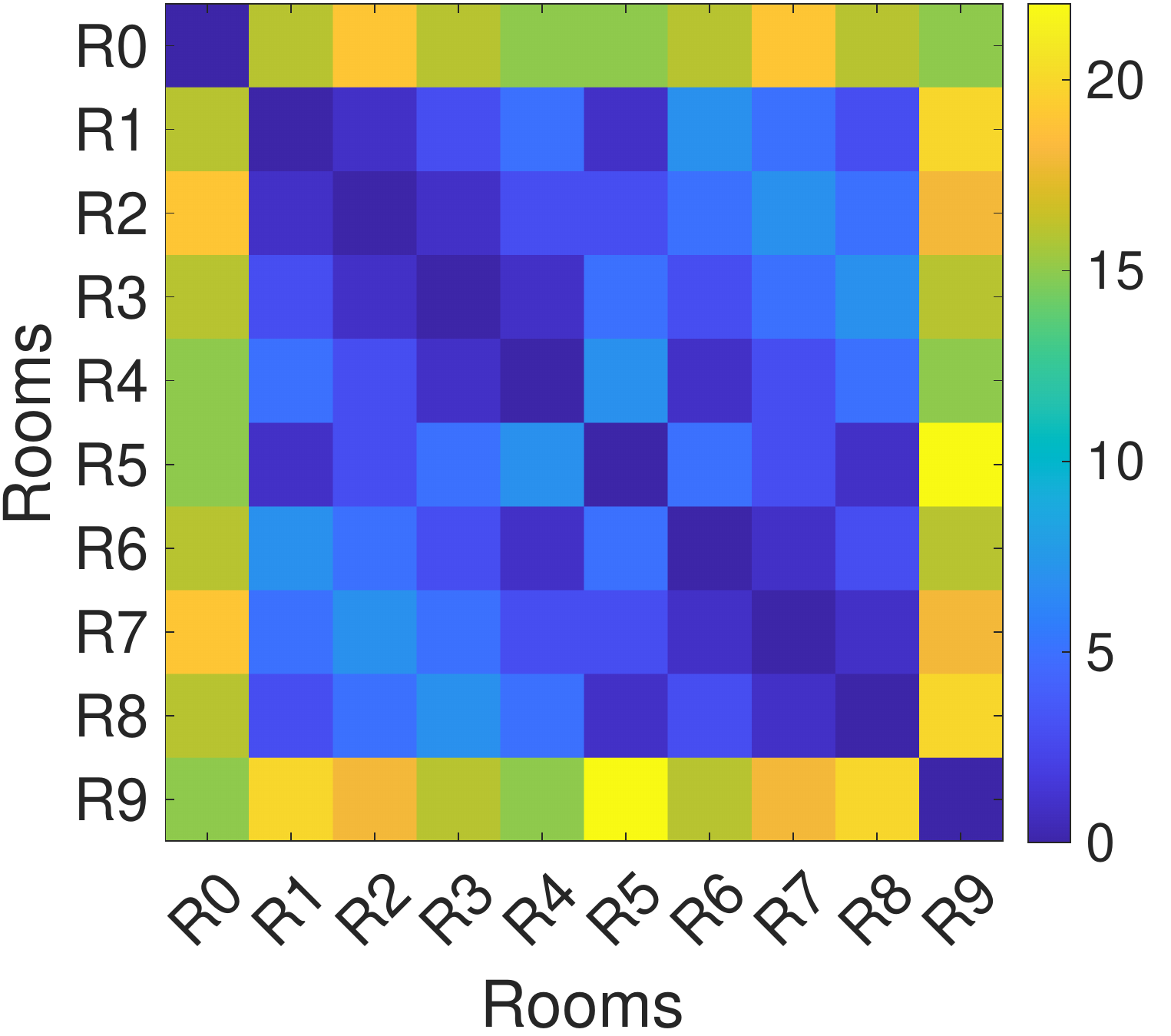}\\
	\textbf{(a)}\hspace{4.8cm}\textbf{(b)}\hspace{5cm}\textbf{(c)}\phantom{xxx}
	\caption{
		\textbf{(a)} Graph representation of Galleria Borghese (cf.\ \fref{fig:RawBeacon}\textbf{(a)} with the corresponding edge-clustering enforced by the orange connections ({\color{orangeM}$\leftrightarrow$}).
		\textbf{(b)} Structure of a cascaded selector based on the clustering in \textbf{(a)}.
		\textbf{(c)} The distance matrix obtained weighting `2' the door connections, `15' the staircases connection and `0' the different type of exhibitions.
		A discount factor of `$\sfrac12$' is applied for the first door transition (cf.\ Footnote \ref{fn:discountFactor}).
		The corresponding numeric values are reported in \aref{app:distanceMatrices} (cf.\ \tref{tab:GBDistanceMatrix}).
	}
	\label{fig:GBgraph}
\end{figure}

In this section, we discuss the performance of our method on the basis of experimental data gathered in our measurement campaign in Galleria Borghese (see \sref{sec:IoT}).
In the following, we first build a total-coloured graph corresponding to the floor plan of GB (cf.\ \fref{fig:RawBeacon}\textbf{(a)}),  and then we design a cascaded localiser relying on the corresponding clustering enforced by the architectural constraints.
Finally, we present the performances of our methods by means of four indicators based on the metric enforced by the graph.

According to the disposition of antennas within the different exhibition areas, we chose to reconstruct the two visiting floors with different resolutions, building trajectories at room-scale on the first floor while having a single comprehensive room on the second one.
We include an entrance/exit artificial room connected to the three entry rooms of the museum ($R_4$, $R_5$ and $R_9$) representing the ticket office (where we also provided visitors with beacons).
From a colour perspective, we consider architectural constraints represented by stairs (cf.\ \fref{fig:GBgraph}\textbf{(a)}).

Building upon the emerging clustering, we develop a two-layer cascaded localiser (cf.\ \fref{fig:GBgraph}\textbf{(b)}) built of two predictors $\mathfrak F$ and $\mathfrak R$ which classify the floor, and the correct room within the sculpture floor F1, respectively.

In \tref{tab:GBresults}, we compare the localisation accuracy we achieve employing the methods presented in \sref{sec:multiLSTM}, considering both online and offline perspectives (i.e.\ involving asymmetric resp.\ symmetric time windows).
Additionally, we report the results obtained via a single classifier (as in~\cite{JCompSci}) as a benchmark.

\begin{table}
	\centering
	\begin{minipage}{0.45\linewidth}
        \scalebox{.575}{
            \renewcommand{\arraystretch}{1.3} 
            \begin{tabular}{| c || c c c c |}
                \multicolumn{5}{c}{Symmetric time window $[-6, +6]$}\\
                \multicolumn{5}{c}{}\\\hline
                Methodology & Accuracy & $\overline{\mathcal D}$ & $\overline{\mathcal D^*}$ & $\overline{\mathcal W}$ \\\hline\hline
                &\multicolumn{4}{c|}{Single Localiser}\\\hline
                ArgMax  & 0.6562 & 4.3755 & 12.727 & 3150.3 \\
                ConvKer & 0.7645 & 3.1693 & 13.458 & 2281.9 \\
                MLP     & 0.8679 & 1.4402 & 10.903 & 1037.0 \\
                LSTM    & \tcbM{yellowYed}{0.9267} & \tcbM{yellowYed}{0.5761} & \tcbM{yellowYed}{7.8605} & \tcbM{yellowYed}{414.84} \\
                RF      & 0.8900 & 1.2150 & 11.046 & 874.84 \\\hline
                &\multicolumn{4}{c|}{Floor Localiser}\\\hline
                ArgMax  & 0.8673 & 1.9905 & 15 & -- \\
                ConvKer & 0.9005 & 1.4925 & 15 & -- \\
                MLP     & 0.9045 & 1.4325 & 15 & -- \\
                LSTM    & \tcbM{yellowYed}{0.9650} & \tcbM{yellowYed}{0.5250} & 15 & -- \\
                RF      & 0.9386 & 0.9210 & 15 & -- \\\hline
                &\multicolumn{4}{c|}{Floor Localiser with downsampling ($\Delta t = 5$min)}\\\hline
                ArgMax  & 0.9148 & 1.2780 & 15 & -- \\
                ConvKer & 0.8201 & 2.6984 & 15 & -- \\
                MLP     & 0.9949 & 0.0764 & 15 & -- \\
                LSTM    & 0.9940 & 0.0900 & 15 & -- \\
                RF      & \tcbM{yellowYed}{0.9966} & \tcbM{yellowYed}{0.0509} & 15 & -- \\\hline
                &\multicolumn{4}{c|}{Sculpture floor Localiser}\\\hline
                ArgMax  & 0.8111 & 0.3314 & 1.7546 & -- \\
                ConvKer & 0.8974 & 0.1327 & 1.2938 & -- \\
                MLP     & 0.9493 & 0.0593 & 1.1714 & -- \\
                LSTM    & \tcbM{yellowYed}{0.9507} & \tcbM{yellowYed}{0.0521} & \tcbM{yellowYed}{1.0588} & -- \\
                RF      & 0.9435 & 0.0709 & 1.2564 & -- \\\hline
                \phantom{LSTM -- LSTM}&\multicolumn{4}{c|}{Multiple Localiser (Floor + Sculpture)}\\\hline
                RF -- LSTM & \tcbM{yellowYed}{\textbf{0.9677}} & \tcbM{yellowYed}{\textbf{0.0816}} & \tcbM{yellowYed}{2.5267} & \tcbM{yellowYed}{\textbf{58.760}} \\\hline
            \end{tabular}
        }
    \end{minipage}
    \hfill
    \begin{minipage}{0.45\linewidth}
        \scalebox{.55}{
            \renewcommand{\arraystretch}{1.3}
            \begin{tabular}{| c || c c c c |}
                \multicolumn{5}{c}{Asymmetric time window $[-6, 0]$}\\
                \multicolumn{5}{c}{}\\\hline
                Methodology & Accuracy & $\overline{\mathcal D}$ & $\overline{\mathcal D^*}$ & $\overline{\mathcal W}$\\\hline\hline
                &\multicolumn{4}{c|}{Single Localiser}\\\hline
                ArgMax  & 0.6562 & 4.3755 & 12.727 & 3150.3 \\
                ConvKer & 0.7105 & 3.4537 & 11.930 & 2486.6 \\
                MLP     & 0.8525 & 1.5483 & 10.497 & 1114.7 \\
                LSTM    & \tcbM{yellowYed}{0.8721} & \tcbM{yellowYed}{1.1459} & \tcbM{yellowYed}{8.9600} & \tcbM{yellowYed}{825.10} \\
                RF      & 0.8662 & 1.3524 & 10.108 & 973.76 \\\hline
                &\multicolumn{4}{c|}{Floor Localiser}\\\hline
                ArgMax  & 0.8673 & 1.9905 & 15 & -- \\
                ConvKer & 0.8986 & 1.5210 & 15 & -- \\
                MLP     & 0.9011 & 0.9480 & 15 & -- \\
                LSTM    & \tcbM{yellowYed}{0.9480} & \tcbM{yellowYed}{0.7800} & 15 & -- \\
                RF      & 0.9258 & 1.1130 & 15 & -- \\\hline
                &\multicolumn{4}{c|}{Floor Localiser with downsampling ($\Delta t = 5$min)}\\\hline
                ArgMax  & 0.9148 & 1.2780 & 15 & -- \\
                ConvKer & 0.8283 & 2.5754 & 15 & -- \\
                MLP     & 0.9932 & 0.1020 & 15 & -- \\
                LSTM    & \tcbM{yellowYed}{0.9966} & \tcbM{yellowYed}{0.0509} & 15 & -- \\
                RF      & 0.9957 & 0.0644 & 15 & -- \\\hline
                &\multicolumn{4}{c|}{Sculpture floor Localiser}\\\hline
                ArgMax  & 0.8111 & 0.3314 & 1.7546 & -- \\
                ConvKer & 0.8258 & 0.2165 & 1.2429 & -- \\
                MLP     & 0.9043 & 0.1072 & 1.1212 & -- \\
                LSTM    & \tcbM{yellowYed}{0.9058} & \tcbM{yellowYed}{0.1260} & 1.3385 & -- \\
                RF      & 0.8928 & 0.1390 & \tcbM{yellowYed}{1.2973} & -- \\\hline
                &\multicolumn{4}{c|}{Multiple Localiser (Floor + Sculpture)}\\\hline
                LSTM -- LSTM & \tcbM{yellowYed}{0.9413} & \tcbM{yellowYed}{0.1250} & 2.1311 & \tcbM{yellowYed}{90.068} \\
                LSTM -- RF   & 0.9337 & 0.1326 & \tcbM{yellowYed}{\textbf{2.0008}} & \tcbM{yellowYed}{95.510} \\\hline
            \end{tabular}
        }
    \end{minipage}
	\caption{
		Localisation performance achieved employing a combination of selectors (\sref{sec:multiLSTM}) on experimental data collected at Galleria Borghese.
		Best results are highlighted in yellow.
		Performances with single localisers are reported as a reference.
		All the localisers have been trained on a desktop computer in less than five minutes, being the RF the fastest in training ($\sfrac{1}{30}$ of MLP training) and $LSTM$ the slowest (three times the MLP training).
		The ConvKer employs a triangular linear kernel.
	}
	\label{tab:GBresults}
\end{table}

On side of the sole accuracy, we include in the comparison other metrics induced by our total-coloured graph and based on the room-distance matrix $\mathcal D$ (cf.\ \fref{fig:GBgraph}\textbf{(c)}, \tref{tab:GBDistanceMatrix}).
We evaluate these metrics on a test set, $\mathcal T$, disjoint from the training set consisting of about $20\%$ of the labelled data.
The metrics considered are
\begin{itemize}
	\item[acc] the ``vanilla'' accuracy, given by the fraction of correct predictions over the total number of samples.
	
	\item[$\overline{\mathcal D}$] mean sample displacement, given by the average distance $\mathcal D$ between the sample, ``instantaneous'', predictions and the corresponding ground truth.
	
	\item[$\overline{\mathcal D}^*$] the mean sample displacement error, given by the average distance $\mathcal D$ restricted to wrongly labelled samples, and the corresponding ground truth.
	
	\item[$\overline{\mathcal W}$] the mean trajectory displacement, given by the average distance $\mathcal W$ between the reconstructed trajectory and the corresponding ground truth.
\end{itemize}
The proposed method proves not only to increase the accuracy but also to commit much smaller errors when a room is misclassified.
In particular, when adopting $\mathfrak F$ = RF and $\mathfrak R$ = LSTM, the mean error committed by the method is $\overline{\mathcal D} \approx 0.08$ compared to $\overline{\mathcal D} \approx 0.57$ by a single LSTM localiser.
This corresponds to an effective displacement restricted to misclassifications of $\overline{\mathcal D^*} \approx 2.5$, where $1$ means the visitor is positioned in an adjacent room and $3$ means the visitor is located 2 rooms away from the ground truth.
We note that this major increase is due to the ``large-scale'' classifier generally labelling correctly (accuracy $>99.5\%$) the floor a visitor is at.
$\mathfrak R$ = LSTM proves to be the best sculpture floor selector, and, overall, using a symmetric time window provides higher results than the online approach.
In the case of the floor selector $\mathfrak F$, online and offline approaches based on RF and LSTM delivered comparable results.
This aspect is probably due to the fact that the floor classification problem is more regular and simpler than the room classification.

\section{Discussion}
\label{sec:conclusions}
In this work, we addressed the issue of increasing the accuracy of room-level tracking of museum visitors based on RSSi analyses.
RSSi-tracking leverages on the measurement of the intensity (RSSi) of self-broadcasting signals emitted by beacons individually assigned to the visitors.
As such, it is a flexible non-invasive method to acquire insights on the guest dynamics with very low hardware requirements, and it is able to overcome most of the museums architectural problems.
However, RSSi are by nature highly fluctuating, requiring more advanced policies than the ``maximum-RSSi'' in order to reconstruct trajectories.

Here, via a graph theory approach, we proposed an encoding for the room-level structure of a museum, including topological information (expert-knowledge) in terms of colours.
Based on the graph encoding we obtained room- and trajectory- metrics and different room-level clustering.
These enable a cascaded localisation approach built of classifiers, organised in layers, that predict the position of a beacon carried by a visitor via progressive refinements of its location.
The advantages are multiple: (i) each predictor may employ a different subset of the antennas, thus lowering the noise of the readings, (ii) lower dimensional feature and output space, hence simpler training procedures, (iii) different classifiers and time-scales which may suit better the distinct classification tasks.
Reaching an accuracy of $>96.7\%$ for our use case in Galleria Borghese, Rome, the proposed approach greatly improves on our previous methodology.
By means of the metric induced by the graph, we get even more insights on the trajectory reconstruction quality, which has a mean displacement error smaller than $0.08$ rooms.

The cascaded structure proposed can be extended to consider multiple progressive clustering as it happens, e.g., for a museum composed of multiple buildings, in which each building is made of multiple floors.
In general, we consider a clustering with $k$ refinements (i.e.\ $k+1$ clustering built hierarchically), we can build a cascaded classifier made of $k+2$ layers of predictors.
Our future work will focus on an automatic approach, based e.g.\ on PCA or SelectK, to identify optimal features and time-scales for each localiser layer.

\subsection*{acknowledgements}

	The authors acknowledge Emiliano Cristiani who envisioned and coordinated the experimental campaign at Galleria Borghese.
	Acknowledged is also the support of Rober\-to Natalini, Sara Suriano, Massimiliano Adamo, Pietro Centorrino, and all the staff of Galleria Borghese.
	EO acknowledges the research project ``SMARTOUR: Intelligent Platform for Tourism'' (No.\ SCN\_00166) funded by the Ministry of University and Research with the Regional Development Fund of European Union (PON Research and Competitiveness 2007-2013).
	AC acknowledges the support of the Talent Scheme (Veni) research programme, through project number 16771, which is financed by the Netherlands Organization for Scientific Research (NWO).

\bibliographystyle{abbrv}
\bibliography{biblio}

\newpage
\appendix

\section{Distance Matrices}\label{app:distanceMatrices}
We report here the distance matrices $\mathcal D$ in the case of the fictitious museum floor plan in \fref{fig:museumExample}\textbf{(a)} and for Galleria Borghese, respectively in \tref{tab:exampleDistanceMatrix} and~\tref{tab:GBDistanceMatrix}.

\begin{table}[h!]
	\centering
	\scalebox{0.5}{
		\begin{tabular}{c||c|cccccccc|cccc}
	\toprule
	& $\tcbM{cyanM}{r_0}$ & $\tcbM{greenM}{r_1}$ & $\tcbM{yellowYed}{r_2}$ & $\tcbM{yellowYed}{r_3}$ & $\tcbM{yellowYed}{r_4}$ & $\tcbM{greenM}{r_5}$ & $\tcbM{yellowYed}{r_6}$ & $\tcbM{yellowYed}{r_7}$ & $\tcbM{yellowYed}{r_8}$ & $\tcbM{greenM}{r_9}$ & $\tcbM{yellowYed}{r_{10}}$ & $\tcbM{yellowYed}{r_{11}}$ & $\tcbM{greenM}{r_{12}}$ \\
	\midrule\midrule
	$\tcbM{cyanM}{r_0}$		    &  0   & 11.5 & 10.5 & 11.5 & 12.5 & 12.5 & 11.5 & 10.5 & 11.5 & 20.5 & 20.5 & 21.5 & 21.5\\\midrule
	$\tcbM{greenM}{r_1}$		& 11.5 &  0   &  1.5 &  2.5 &  3.5 &  1   &  2.5 &  3.5 &  4.5 & 10   & 10.5 & 11.5 & 11 \\
	$\tcbM{yellowYed}{r_2}$		& 10.5 &  1.5 &  0   &  1   &  2   &  2.5 &  3   &  4   &  3   & 10.5 & 10   & 11   & 11.5 \\
	$\tcbM{yellowYed}{r_3}$		& 11.5 &  2.5 &  1   &  0   &  1   &  3.5 &  4   &  3   &  2   & 11.5 & 11   & 12   & 12.5 \\
	$\tcbM{yellowYed}{r_4}$		& 12.5 &  3.5 &  2   &  1   &  0   &  4.5 &  3   &  2   &  1   & 12.5 & 12   & 13   & 13.5 \\
	$\tcbM{greenM}{r_5}$		& 12.5 &  1   &  2.5 &  3.5 &  4.5 &  0   &  1.5 &  2.5 &  3.5 & 11   & 11.5 & 12.5 & 12   \\
	$\tcbM{yellowYed}{r_6}$		& 11.5 &  2.5 &  3   &  4   &  3   &  1.5 &  0   &  1   &  2   & 12.5 & 12   & 13   & 13.5 \\
	$\tcbM{yellowYed}{r_7}$		& 10.5 &  3.5 &  4   &  3   &  2   &  2.5 &  1   &  0   &  1   & 13.5 & 13   & 14   & 14.5 \\
	$\tcbM{yellowYed}{r_8}$		& 11.5 &  4.5 &  3   &  2   &  1   &  3.5 &  2   &  1   &  0   & 13.5 & 13   & 14   & 14.5 \\\midrule
	$\tcbM{greenM}{r_9}$		& 20.5 & 10   & 10.5 & 11.5 & 12.5 & 11   & 12.5 & 13.5 & 13.5 &  0   &  1.5 &  2.5 &  1   \\
	$\tcbM{yellowYed}{r_{10}}$	& 20.5 & 10.5 & 10   & 11   & 12   & 11.5 & 12   & 13   & 13   &  1.5 &  0   &  1   &  2.5 \\
	$\tcbM{yellowYed}{r_{11}}$	& 21.5 & 11.5 & 11   & 12   & 13   & 12.5 & 13   & 14   & 14   &  2.5 &  1   &  0   &  1.5 \\
	$\tcbM{greenM}{r_{12}}$	    & 21.5 & 11   & 11.5 & 12.5 & 13.5 & 12   & 13.5 & 14.5 & 14.5 &  1   &  2.5 &  1.5 &  0   \\
	\bottomrule
\end{tabular}
	}
	\caption{
		Distance matrix $\mathcal D$ from the total-coloured graph in \fref{fig:museumExample}\textbf{(a)}.
		We weight `1' the door connections ($\leftrightarrow$), `10' the staircase links (${\color{orangeM}\leftrightarrow}$) and $+0.5$ the distance between two rooms not sharing the same room-colour.
	}
	\label{tab:exampleDistanceMatrix}
\end{table}

\begin{table}[h!]
	\centering
	\scalebox{0.5}{
		\begin{tabular}{c||c|cccccccc|c}
	\toprule
	& $\tcbM{cyanM}{r_0}$ & $\tcbM{yellowYed}{r_1}$ & $\tcbM{yellowYed}{r_2}$ & $\tcbM{yellowYed}{r_3}$ & $\tcbM{yellowYed}{r_4}$ & $\tcbM{yellowYed}{r_5}$ & $\tcbM{yellowYed}{r_6}$ & $\tcbM{yellowYed}{r_7}$ & $\tcbM{yellowYed}{r_8}$ & $\tcbM{greenM}{r_9}$ \\
	\midrule\midrule
	$\tcbM{cyanM}{r_0}$		    &  0 & 16 & 19 & 16 & 15 & 15 & 16 & 19 & 16 & 15 \\\midrule
	$\tcbM{yellowYed}{r_1}$		& 16 &  0 &  1 &  3 &  5 &  1 &  7 &  5 &  3 & 20 \\
	$\tcbM{yellowYed}{r_2}$		& 19 &  1 &  0 &  1 &  3 &  3 &  5 &  7 &  5 & 18 \\
	$\tcbM{yellowYed}{r_3}$		& 16 &  3 &  1 &  0 &  1 &  5 &  3 &  5 &  7 & 16 \\
	$\tcbM{yellowYed}{r_4}$		& 15 &  5 &  3 &  1 &  0 &  7 &  1 &  3 &  5 & 15 \\
	$\tcbM{yellowYed}{r_5}$		& 15 &  1 &  3 &  5 &  7 &  0 &  5 &  3 &  1 & 22 \\
	$\tcbM{yellowYed}{r_6}$		& 16 &  7 &  5 &  3 &  1 &  5 &  0 &  1 &  3 & 16 \\
	$\tcbM{yellowYed}{r_7}$		& 19 &  5 &  7 &  5 &  3 &  3 &  1 &  0 &  1 & 18 \\
	$\tcbM{yellowYed}{r_8}$		& 16 &  3 &  5 &  7 &  5 &  1 &  3 &  1 &  0 & 20 \\\midrule
	$\tcbM{greenM}{r_9}$		& 15 & 20 & 18 & 16 & 15 & 22 & 16 & 18 & 20 &  0 \\
	\bottomrule
\end{tabular}

	}
	\caption{
		Distance matrix $\mathcal D$ from GB total-coloured graph in \fref{fig:GBgraph}\textbf{(a)}.
		We weight `2' the door connections ($\leftrightarrow$), `15' the staircase links (${\color{orangeM}\leftrightarrow}$) and we apply a discount of $\sfrac12$ on the first door transition (cf.\ \fref{fig:GBgraph}\textbf{(c)} for  heat map corresponding to $\mathcal D$).
	}
	\label{tab:GBDistanceMatrix}
\end{table}

\end{document}